\documentclass[prl, twocolumn, groupedaddress]{revtex4-1}%
\usepackage{amsfonts}
\usepackage{amsmath}
\usepackage{amssymb}
\usepackage{graphicx}
\usepackage{bm}
\usepackage{subfigure}
\usepackage{color}
\usepackage[colorlinks=true,linktocpage=true,breaklinks=true]{hyperref}%
\providecommand{\U}[1]{\protect\rule{.1in}{.1in}}
\begin{document}
\title{Positive longitudinal spin magnetoconductivity in $\mathbb{Z}_{2}$ topological
Dirac semimetals }
\author{Ming-Xun Deng$^{1}$}
\author{Yan-Yan Yang$^{1}$}
\author{Wei Luo$^{2}$}
\author{R. Ma$^{3}$}
\author{Rui-Qiang Wang$^{1}$}
\email{wangruiqiang@m.scnu.edu.cn}
\author{L. Sheng$^{4,5}$}
\email{shengli@nju.edu.cn}
\author{D. Y. Xing$^{4,5}$}
\affiliation{$^{1}$Guangdong Provincial Key Laboratory of Quantum Engineering and Quantum
Materials, GPETR Center for Quantum Precision Measurement, SPTE, South China
Normal University, Guangzhou 510006, China\\
$^{2}$School of Science, Jiangxi University of Science and Technology,
Ganzhou 341000, China\\
$^{3}$Jiangsu Key Laboratory for Optoelectronic Detection of Atmosphere and
Ocean, Nanjing University of Information Science and Technology, Nanjing
210044, China\\
$^{4}$National Laboratory of Solid State Microstructures and Department of
Physics, Nanjing University, Nanjing 210093, China\\
$^{5}$Collaborative Innovation Center of Advanced Microstructures, Nanjing
University, Nanjing 210093, China }

\begin{abstract}
Recently, a class of Dirac semimetals, such as \textrm{Na}$_{\mathrm{3}}%
$\textrm{Bi} and \textrm{Cd}$_{\mathrm{2}}$\textrm{As}$_{\mathrm{3}}$, are
discovered to carry $\mathbb{Z}_{2}$ monopole charges. We present an
experimental mechanism to realize the $\mathbb{Z}_{2}$ anomaly in regard to
the $\mathbb{Z}_{2}$ topological charges, and propose to probe it by
magnetotransport measurement. In analogy to the chiral anomaly in a Weyl
semimetal, the acceleration of electrons by a spin bias along the magnetic
field can create a $\mathbb{Z}_{2}$ charge imbalance between the Dirac points,
the relaxation of which contributes a measurable positive longitudinal spin
magnetoconductivity (LSMC) to the system. The $\mathbb{Z}_{2}$ anomaly induced
LSMC is a spin version of the longitudinal magnetoconductivity (LMC) due to
the chiral anomaly, which possesses all characters of the chiral anomaly
induced LMC. While the chiral anomaly in the topological Dirac semimetal is
very sensitive to local magnetic impurities, the $\mathbb{Z}_{2}$ anomaly is
found to be immune to local magnetic disorder. It is further demonstrated that the
quadratic or linear field dependence of the positive LMC is not
unique to the chiral anomaly. Base on this, we argue that the
periodic-in-$1/B$ quantum oscillations superposed on the positive LSMC can serve
as a fingerprint of the $\mathbb{Z}_{2}$ anomaly in topological Dirac semimetals.

\end{abstract}
\maketitle

Topological semimetals are novel quantum states of matter, where the conduction and
valence bands touch, near the Fermi level, at certain discrete momentum
points or
lines~\cite{RevModPhys.90.015001,Liu864,Zhang:2016aa,Xiong413,Zhang:2017aa,PhysRevLett.119.136806}.
The gap-closing points or lines are protected either by crystalline symmetry
or topological invariants~\cite{Yang:2014aa,PhysRevB.91.121101,Kargarian8648}.
A topological Dirac semimetal hosts stable gap-closing points called the Dirac
points (DPs), which, in addition to the time-reversal (TR) and
spatial-inversion (SI) symmetries, are protected by the crystalline symmetry.
By breaking the TR or SI symmetry, a single DP can split into a pair of Weyl
nodes, leading to the topological transition from a Dirac to a Weyl
semimetal~\cite{PhysRevX.9.011039,PhysRevB.85.165110,PhysRevB.88.245107,PhysRevB.98.085149,PhysRevB.96.155141,PhysRevB.99.075131}. The Weyl nodes always come in pairs with opposite chiralities in momentum
space, protected by topological invariants associated with the Chern flux and
connected by the nonclosed Fermi-arc surface
states~\cite{NIELSEN1983389,Volovik2003,PhysRevB.83.205101}.

The Fermi-arc surface states are regarded as the most distinctive observable
spectroscopic feature of Weyl semimetals. However, their observation is
sometime limited by spectroscopic resolutions. Therefore, there is an urgency
to find similar smoking-gun features of Weyl semimetals in response,
especially in transport measurements. Of particular interest is the transport related to
the chiral anomaly, i.e., the violation of the separate number conservation
laws of Weyl fermions of different chiralities. Nonorthogonal electric and
magnetic fields can pump Weyl fermions between Weyl nodes of opposite
chiralities, and create a population imbalance between them. The relaxation of
the chirality population imbalance contributes an extra electric current to
the system, which results in a very unusual positive longitudinal
magnetoconductivity
(LMC)~\cite{PhysRevX.5.031023,PhysRevLett.122.036601,PhysRevX.8.031002}. While
the positive LMC, as a condensed-matter manifestation of the chiral anomaly,
was observed recently in Weyl semimetal~\textrm{TaAs}~
\cite{PhysRevX.5.031023,Zhang:2016aa}. It was also observed in Dirac
semimetals~\textrm{Na}$_{\mathrm{3}}$\textrm{Bi}\cite{Liu864,Xiong413} and
\textrm{Cd}$_{\mathrm{2}}$\textrm{As}$_{\mathrm{3}}$%
\cite{Neupane:2014aa,Li:2015aa,Zhang:2017aa}. It is now understood that
\textrm{Na}$_{\mathrm{3}}$\textrm{Bi} and \textrm{Cd}$_{\mathrm{2}}%
$\textrm{As}$_{\mathrm{3}}$, protected by a nontrivial $\mathbb{Z}_{2}$
topological invariant, belong to a new class of Dirac semimetals, in which the
DPs occur in pairs and separate in momentum space along a rotation
axis~\cite{Kargarian8648,Yang:2014aa,PhysRevB.91.121101}. The momenta of the
Dirac fermions in these Dirac semimetals are locked to their spin and orbital
parity, simultaneously. The Weyl nodes at the same DP belonging to different
irreducible representations in spin subspace cannot be coupled, and
have to seek for a partner with the same spin from the other DP. As a
consequence, the two DPs composing of two pairs of Weyl nodes are connected by
two Fermi arcs~\cite{,A3B_PhysRevB.85.195320,Cd3As2_PhysRevB.88.125427}, much
like in the Weyl semimetals~\cite{PhysRevB.83.205101}.

Naturally, one may ask, in analogy to the chiral anomaly, whether there exists
$\mathbb{Z}_{2}$ anomaly in regard to the $\mathbb{Z}_{2}$ topological charge.
If there exists, how it manifests in experiments, or how to identify the
$\mathbb{Z}_{2}$ anomaly? In Ref. \cite{PhysRevLett.117.136602}, by
introducing a fictitious spin gauge field which couples antisymmetrically to
the spin, Burkov and Kim answered the first question in the affirmative. In
this paper, we present an experimental mechanism to realize the
$\mathbb{Z}_{2}$ anomaly for Dirac semimetals carrying the $\mathbb{Z}_{2}$
monopole charges, such as \textrm{Na}$_{\mathrm{3}}$\textrm{Bi} and
\textrm{Cd}$_{\mathrm{2}}$\textrm{As}$_{\mathrm{3}}$, and then we propose to
probe the $\mathbb{Z}_{2}$ anomaly by magnetotransport measurement. As we
show, the $\mathbb{Z}_{2}$ anomaly, in fact, is a spin version of the chiral
anomaly, in which the acceleration of electrons by a spin bias along the
magnetic field can create carrier density imbalance between the DPs, the
relaxation of which leads to a measurable positive longitudinal spin
magnetoconductivity (LSMC). We further demonstrate that the $B^{2}$ or $B$
dependence emerging in the positive LMC are not unique to the chiral anomaly.
Like the quantum oscillations of the positive LMC in Weyl
semimetals~\cite{PhysRevLett.117.077201,PhysRevLett.122.036601}, we argue that
the periodic-in-$1/B$ quantum oscillations superposed on the positive LSMC are
remarkable fingerprint of the $\mathbb{Z}_{2}$ anomaly in topological Dirac semimetals.

\begin{figure}[ptb]
\centering
\includegraphics[width=\linewidth]{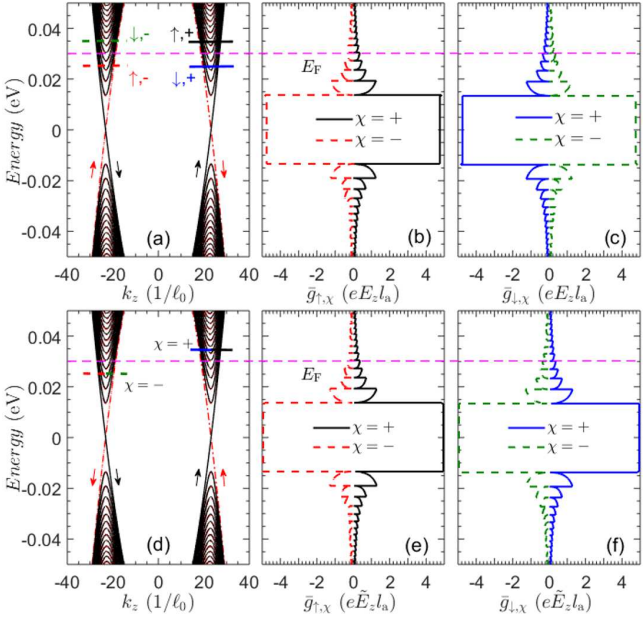}\caption{The nonequilibium local
chemical potential induced by the external fields, with the upper panel for
$\mathbf{\nabla}\mu_{s}=-eE_{z}\hat{e}_{z}$ and the lower panel for
$\mathbf{\nabla}\mu_{s}=-es\tilde{E}_{z}\hat{e}_{z}$. The first column
displays the LLs for spin-$\uparrow$ (dark-solid) and spin-$\downarrow$
(red-dash-dotted) Weyl fermions, where the parallel short lines are an
enlarged illustration of the nonequilibium local chemical potential of each
Weyl valleys due to the effect of chiral and $\mathbb{Z}_{2}$ anomalies. The
arrows denote the driven directions of the Weyl fermions in the dispersion by
the external fields and $1/\ell_{0}=\sqrt{\frac{e(B=1T)}{\hbar}}$ is unit of
momentum. The last two columns, { carried out for }\textrm{Na}$_{3}%
$\textrm{Bi}\cite{A3B_PhysRevB.85.195320}, are numerical solutions of Eq.
{(\ref{eq_gfun}), where }$l_{\mathrm{a}}\equiv\tilde{\upsilon}_{\mathrm{F}%
}\tau_{\mathrm{intra}}$ represents the intravalley relaxation length.{ }Other
parameters are chosen as{ }$B=1T$, $\tau_{\mathrm{inter}}^{\mathrm{c}}%
=5\tau_{\mathrm{intra}}$,{ }$\tau_{\mathrm{intra}}^{\mathrm{s}}=20\tau
_{\mathrm{inter}}^{\mathrm{c}}$ and $\tau_{\mathrm{inter}}^{\mathrm{s}%
}=100\tau_{\mathrm{inter}}^{\mathrm{c}}$. }%
\label{figLLs}%
\end{figure}

We start from the general low-energy Hamiltonian for topological Dirac
semimetals \textrm{Na}$_{3}$\textrm{Bi}~\cite{A3B_PhysRevB.85.195320}
and\textrm{ Cd}$_{3}\mathrm{As}_{2}$~\cite{Cd3As2_PhysRevB.88.125427}
\begin{equation}
H(\mathbf{k})=\hbar\upsilon_{\mathrm{F}}(\sigma^{x}s^{z}k_{x}-\sigma^{y}%
k_{y})+m(\mathbf{k})\sigma^{z}+\mathcal{O}(|\mathbf{k}|^{2}) \label{eq_Hk}%
\end{equation}
with $m(\mathbf{k})=(m_{0}-m_{1}k_{z}^{2})-m_{2}(k_{x}^{2}+k_{y}^{2})$, where
$s^{x,y,z}$ and $\sigma^{x,y,z}$ are Pauli matrices acting on the spin and
orbital parity degrees of freedom, respectively. $\mathcal{O}(|\mathbf{k}%
|^{2})$ is a higher-order term in momentum related to the rotational
symmetries of the crystal structures, which, in the vicinity of the
gap-closing points, is negligible. Therefore, $[s^{z},H(\mathbf{k})]=0$, and
the Hamiltonian separates into two independent $2\times2$ blocks, which can be
labelled by the eigenvalues of $s^{z}$, namely, $s=\pm1$. Each spin block contributes
a Weyl node at the DPs $\mathbf{k}_{w}^{\chi}=(0,0,\chi\sqrt{m_{0}/m_{1}})$,
where $\chi=\pm1$ refer to the $\mathbb{Z}_{2}$ charges of the DPs.

Consider the topological Dirac semimetal subjected to an electromagnetic
field, which can be described by the Hamiltonian $H\left(  \mathbf{k}%
+e\mathbf{A}(\mathbf{r})/\hbar\right)  $, where $\mathbf{A}(\mathbf{r})$ is a
vector potential for the electromagnetic field. In a uniform magnetic field
applied along the $z$ direction, i.e., $\mathbf{A}(\mathbf{r})=Bx\hat{e}_{y}$,
the energy spectrum can be solved exactly, yielding $E_{n,k_{z}}^{s}%
=-sm_{2}/\ell_{B}^{2}+\varepsilon_{n,k_{z}}^{s}$ with%
\begin{equation}
\varepsilon_{n,k_{z}}^{s}=%
\begin{cases}
s(\Lambda_{0}-m_{1}k_{z}^{2}) & n=0\\
\mathrm{sgn}(n)\sqrt{(\Lambda_{n}-m_{1}k_{z}^{2})^{2}+2|n|(\hbar\omega
_{c})^{2}} & n\neq0
\end{cases}
, \label{eq_LLs}%
\end{equation}
where $\Lambda_{n}=m_{0}-2|n|m_{2}/\ell_{B}^{2}$, $\omega_{c}=\upsilon
_{\mathrm{F}}/\ell_{B}$ and $\ell_{B}=\sqrt{\hbar/eB}$. The Landau levels
(LLs) are plotted in Figs. \ref{figLLs}(a) and (d), each of which has a
degeneracy equal to $1/2\pi\ell_{B}^{2}$ per unit cross section. Notice that,
due to the coupling between the magnetic field and the electron orbital
angular momentum, a spin-dependent term $-sm_{2}/\ell_{B}^{2}$ appears in the
spectrum, which shifts the energies of the Weyl fermions
of opposite spins in opposite directions,
and thus lifts the spin degeneracy.

As we focus on the physics around the gap-closing points, it is convenient to
expand Eq. (\ref{eq_LLs}) near the Weyl nodes. To linear order, we obtain%
\begin{equation}
\varepsilon_{s,n}^{\chi}(q_{z})=\mbox{sgn}(n)\hbar\sqrt{2|n|\omega_{c}%
^{2}+\tilde{\upsilon}_{\mathrm{F}}^{2}q_{z}^{2}}+s\chi\hbar\tilde{\upsilon
}_{\mathrm{F}}q_{z}\delta_{n,0}, \label{eq_lls}%
\end{equation}
where $\tilde{\upsilon}_{\mathrm{F}}=2\hbar^{-1}\sqrt{m_{0}m_{1}}$ and
$\mathbf{q}=\mathbf{k}-\mathbf{k}_{w}^{\chi}$ is momentum measured from the
Weyl nodes. As it shows, in each Weyl valley, the $n=0$ LL is chiral,
manifesting the chirality of the Weyl node, and all $n\neq0$ LLs are achiral.
In the presence of an electric field, the system will exhibit the chiral
anomaly\cite{PhysRevB.86.115133,PhysRevLett.120.026601}, i.e., the
acceleration of the fermions by the electric field creates a chirality
population imbalance between the Weyl valleys, and then leads to a measurable
positive LMC. Here, the chiral $n=0$ LLs are not only $\mathbb{Z}_{2}$- but
also spin-resolved. Moreover, for a single pair of Weyl nodes for a fixed
spin $s$, the $\mathbb{Z}_{2}$ charge $\chi$ in Eq. (\ref{eq_lls}) plays the
role of the chirality, which exhibits the $\mathbb{Z}_{2}$ quantum anomaly.
The chirality manifested in the $n=0$ LL, in fact, can be understood as
follows. The $\mathbb{Z}_{2}$ charge of the DP is defined as $\chi
=(C_{\uparrow}^{\chi}-C_{\downarrow}^{\chi})/2$, where $C_{\uparrow
,\downarrow}^{\chi}$ are the chiralities of the spin-$\uparrow$ and
spin-$\downarrow$ Weyl fermions at the $\chi$
DP~\cite{Yang:2014aa,PhysRevB.91.121101}. The paired Weyl nodes possess
opposite chiralities $C_{s}^{\chi}=-C_{s}^{-\chi}$, and therefore, the
chirality of each Weyl node here is $C_{s}^{\chi}=s\chi$, which is exactly the
sign of the $n=0$ LL's group velocity, as shown in Eq.\ (\ref{eq_lls}).
Recalling the mechanism of the positive LMC in Weyl
semimetals~\cite{PhysRevLett.122.036601}, the $\mathbb{Z}_{2}$ anomaly may also
contribute a measurable physical quantity, which is similar to the chiral
anomaly in response to the parallel electric and magnetic
fields.\begin{figure}[ptb]
\centering
\includegraphics[width=\linewidth]{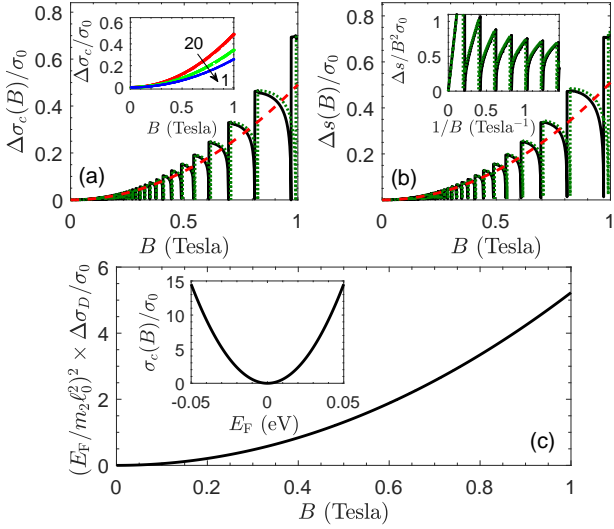}\caption{(a) $\Delta\sigma
_{c}(B)$, (b) $\Delta s(B)$ and (c) $\Delta\sigma_{D}$ vs the magnetic field
$B$. The dark-solid and cyan-dotted lines in (a) and (b) stand for the
spin-$\uparrow$ and spin-$\downarrow$ components, where the red-dashed lines
indicate the envelopes of the {oscillations, }described by the classical
formula in Refs. \cite{PhysRevB.88.104412,PhysRevB.96.165101}. The solid lines
in the inset of (a) shows the envelopes of the LMC for defferent magnetic
doping concentration, with red, green and blue for $\tau_{\mathrm{intra}%
}^{\mathrm{s}}/\tau_{\mathrm{inter}}^{\mathrm{c}}=(20,2,1)$. The data of (b)
are replotted in its inset to show the periodic-in-$1/B$ dependence{ of
}$\Delta s(B)$, and the inset of (c) displays the {Drude conductivity as a
function of the Fermi level. Here, }$E_{\mathrm{F}}=0.03$ \textrm{eV
}corresponding to the pink dashed lines in Fig. \ref{figLLs}, $\tau
_{\mathrm{intra}}^{\mathrm{s}}/\tau_{\mathrm{inter}}^{\mathrm{c}}=20$, and
$\sigma_{0}=(e^{2}/h)\tilde{\upsilon}_{\mathrm{F}}\tau_{\mathrm{intra}}%
/\ell_{0}^{2}$ is chosen as the unit of conductivity for convenience.}%
\label{figcon}%
\end{figure}

To demonstrate this effect, let us couple an external field, e.g., an electric
field $\mathbf{\nabla}\mu_{s}=-e\mathbf{E}$ or a spin-dependent electric field
$\mathbf{\nabla}\mu_{s}=-es\tilde{\mathbf{E}}$ which can be induced by a spin
bias, to the fermions. Upon application of the external field, the
linear-response electron distribution function in general takes the form%
\begin{equation}
f_{s,n}^{\chi}(\mathbf{k})=f_{0}(E_{n,k_{z}}^{s})+\left[  -\frac{\partial
f_{0}(\varepsilon_{s,n}^{\chi})}{\partial\varepsilon_{s,n}^{\chi}}\right]
g_{s,n}^{\chi}(\mathbf{k}), \label{eq_fsn}%
\end{equation}
where $g_{s,n}^{\chi}(\mathbf{k})$ describes the deviation of $f_{s,n}^{\chi
}(\mathbf{k})$ from the electron {equilibrium }distribution function
$f_{0}(\varepsilon_{s,n}^{\chi})=1/\left[  1+\exp(\frac{\varepsilon
_{s,n}^{\chi}-E_{\mathrm{F}}^{s}}{k_{B}T})\right]  $, with $E_{\mathrm{F}}%
^{s}=E_{\mathrm{F}}+sm_{2}/\ell_{B}^{2}$. {In }the relaxation time
approximation, the steady-state Boltzmann equation for the multiple Fermi
pocket system can be expressed
as~\cite{PhysRevB.34.2147,PhysRevB.89.195137,PhysRevB.99.085405,PhysRevB.99.165146}%
\begin{equation}
\boldsymbol{\upsilon}_{s,n}^{\chi}\cdot\mathbf{\nabla}\mu_{s}=\sum_{s^{\prime
}\chi^{\prime}}\frac{g_{s,n}^{\chi}-\bar{g}_{ss^{\prime}}^{\chi\chi^{\prime}}%
}{\tau_{s,s^{\prime}}^{\chi,\chi^{\prime}}}, \label{eq_gfun}%
\end{equation}
where $\bar{g}_{ss^{\prime}}^{\chi\chi^{\prime}}=(\bar{g}_{s,\chi}+\bar
{g}_{s^{\prime},\chi^{\prime}})/2$ denotes the {equilibrium established
between the }Fermi pockets and $\tau_{s,s^{\prime}}^{\chi,\chi^{\prime}}$
represents the relaxation time due to disorder. The group velocities for the
LLs, given by $\hbar\boldsymbol{\upsilon}_{s,n}^{\chi}=\mathbf{\nabla
}_{\mathbf{q}}\varepsilon_{s,n}^{\chi}(q_{z})$, correspond to the slopes of
the dispersion and the average $\bar{g}_{s,\chi}=\langle g_{s,n}^{\chi
}(\mathbf{k})\rangle_{s,\chi}$ is defined as
\begin{equation}
\langle\cdots\rangle_{s,\chi}=\frac{\sum_{n,q_{y}}\int dq_{z}(\cdots)\left[
-\frac{\partial f_{0}(E_{n,k_{z}}^{s})}{\partial\varepsilon_{s,n}^{\chi}%
}\right]  }{\sum_{n,q_{y}}\int dq_{z}\left[  -\frac{\partial f_{0}(E_{n,k_{z}%
}^{s})}{\partial\varepsilon_{s,n}^{\chi}}\right]  }, \label{eq_ave}%
\end{equation}
where the summation runs over all electron states at the Fermi pocket in the
$\chi$ valley of $s$ spin {component}. It is assumed $\tau_{s,s}^{\chi,-\chi
},\tau_{s,-s}^{\chi,\chi},\tau_{s,-s}^{\chi,-\chi}\gg\tau_{s,s}^{\chi,\chi}$,
based on the fact that, on one hand, the separation of the Weyl nodes usually
makes the intervalley scattering much weaker than intravalley scattering, and
on the other hand, the $z$ component of the spin, served as a conserved
quantity, will have a long relaxation time for dilute magnetically doping. For
the sake of brevity, we denote $\tau_{\mathrm{intra}}=\tau_{s,s}^{\chi,\chi}$
($\tau_{\mathrm{intra}}^{\mathrm{s}}=\tau_{s,-s}^{\chi,\chi}$) and
$\tau_{\mathrm{inter}}^{\mathrm{c}}=\tau_{s,s}^{\chi,-\chi}$ ( $\tau
_{\mathrm{inter}}^{\mathrm{s}}=\tau_{s,-s}^{\chi,-\chi}$) as
intra-Dirac-valley and inter-Dirac-valley relaxation times due to charged
(magnetic) impurity scattering.

\emph{Electric field induced chiral chemical potential- }For $\mathbf{\nabla
}\mu_{s}=-e\mathbf{E}$, the fermions in the two spin components are
accelerated by the electric field toward the same direction. Since the chiral
$n=0$ LLs depend not only on the $\mathbb{Z}_{2}$ charge $\chi$ but also on
the spin $s$, the spin-$\uparrow$ Weyl fermions are pumped from the negative
to the positive chirality, while it reverses for the spin-$\downarrow$ Weyl
fermions, indicated by the dark and red arrows in Fig. \ref{figLLs}(a). As a
result, the global{ equilibrium can be }established by electron scattering
between Weyl valleys residing at distinct or identical DPs, which includes two
different relaxation processes: (i) identical spin component but different
$\mathbb{Z}_{2}$ charges and (ii) identical $\mathbb{Z}_{2}$ charge but
different spin components. In this case, we can reduce Eq. {(\ref{eq_gfun}) to
}%
\begin{equation}
e\mathbf{E}\cdot\boldsymbol{\upsilon}_{s,n}^{\chi}=-\frac{g_{s,n}^{\chi}%
-\bar{g}_{s,\chi}}{\tau_{\mathrm{intra}}}-\frac{g_{s,n}^{\chi}-\bar{g}_{s}%
}{\tau_{\mathrm{inter}}^{\mathrm{c}}}-\frac{g_{s,n}^{\chi}-\bar{g}_{c}}%
{\tau_{\mathrm{intra}}^{\mathrm{s}}} \label{eq_gsnE}%
\end{equation}
with $\bar{g}_{s}=(\bar{g}_{s,\chi}+\bar{g}_{s,-\chi})/2$ and $\bar{g}%
_{c}=(\bar{g}_{s,\chi}+\bar{g}_{-s,\chi})/2$. For $\tau_{\mathrm{inter}%
}^{\mathrm{c}},\tau_{\mathrm{intra}}^{\mathrm{s}}\gg\tau_{\mathrm{intra}}$,
{we can} approximate $g_{s,n}^{\chi}\simeq\bar{g}_{s,\chi}$ in the last two
terms of Eq. {(\ref{eq_gsnE}), and thus arrive at }$g_{s,n}^{\chi
}=-e\mathbf{E}\cdot\boldsymbol{\upsilon}_{s,n}^{\chi}\tau_{\mathrm{intra}%
}+\bar{g}_{s,\chi}$ with{ }
\begin{equation}
\bar{g}_{s,\chi}=-e\mathbf{E}\cdot\left\langle \boldsymbol{\upsilon}%
_{s,n}^{\chi}\right\rangle _{s,\chi}\frac{\tau_{\mathrm{inter}}^{\mathrm{c}%
}\tau_{\mathrm{intra}}^{\mathrm{s}}}{\tau_{\mathrm{inter}}^{\mathrm{c}}%
+\tau_{\mathrm{intra}}^{\mathrm{s}}}, \label{eq_dus}%
\end{equation}
where $m_{2}$, $\tau_{\mathrm{intra}}/\tau_{\mathrm{inter}}^{\mathrm{c}}$ and
$\tau_{\mathrm{intra}}/\tau_{\mathrm{intra}}^{\mathrm{s}}$ are dropped first
due to smallness. It is noticed that only the chiral $n=0$ LLs make a nonzero
contribution to $\left\langle \boldsymbol{\upsilon}_{s,n}^{\chi}\right\rangle
_{s,\chi}$ and, in turn, to $\bar{g}_{s,\chi}$. As $\boldsymbol{\upsilon
}_{s,0}^{\chi}\propto s\chi$, the sign of $\left\langle \boldsymbol{\upsilon
}_{s,n}^{\chi}\right\rangle _{s,\chi}$ is determined by the product of $s$ and
$\chi$. According to Eq.\ {(\ref{eq_fsn}), }$\bar{g}_{s,\chi}$ in fact
corresponds to the nonequilibium local chemical potential in the Weyl valleys.
Therefore, we define the chiral chemical potential for each spin component as
$\Delta\mu_{s}=(\bar{g}_{s,+}-\bar{g}_{s,-})/2$. The chiral chemical
potentials for the two spin components are equal in magnitude but opposite in
the signs, as shown in Figs. \ref{figLLs}(a)-(c). For dilute magnetically
doping $\tau_{\mathrm{intra}}^{\mathrm{s}}\gg\tau_{\mathrm{inter}}%
^{\mathrm{c}}$, $\Delta\mu_{s}=$ $-e\chi\mathbf{E}\cdot\left\langle
\boldsymbol{\upsilon}_{s,n}^{\chi}\right\rangle _{s,\chi}\tau_{\mathrm{inter}%
}^{\mathrm{c}}$, which recovers the result for Weyl semimetals of a single
pair of nodes~\cite{PhysRevLett.122.036601}. Here, as the magnetic impurity
scattering strengthens, the chiral chemical potential will reduce quickly, as
indicated by Eq.\ {(\ref{eq_dus})}. With further increasing the magnetic doping
concentration, $\tau_{\mathrm{intra}}^{\mathrm{s}}<\tau_{\mathrm{inter}%
}^{\mathrm{c}}$ could be accessible, and then the chiral chemical potentials
turns to be very sensitive to the local magnetic disorder.

\emph{Spin bias induced }$\mathbb{Z}_{2}$\emph{ chemical potential- }For
$\mathbf{\nabla}\mu_{s}=-es\tilde{\mathbf{E}}$, the Weyl fermions in the two
spin components are accelerated toward opposite directions, such that the
global{ equilibrium can only be established by electron scattering between
different DPs. In this situation, }Eq. {(\ref{eq_gfun}) reduces to be }%
\begin{equation}
es\tilde{\mathbf{E}}\cdot\boldsymbol{\upsilon}_{s,n}^{\chi}=-\frac
{g_{s,n}^{\chi}-\bar{g}_{s,\chi}}{\tau_{\mathrm{intra}}}-\frac{g_{s,n}^{\chi
}-\bar{g}_{s}}{\tau_{\mathrm{inter}}^{\mathrm{c}}}-\frac{g_{s,n}^{\chi}%
-\bar{g}_{z}}{\tau_{\mathrm{inter}}^{\mathrm{s}}}\ \label{eq_gsnS}%
\end{equation}
with $\bar{g}_{z}=(\bar{g}_{s,\chi}+\bar{g}_{-s,-\chi})/2$. From Eq.
{(\ref{eq_gsnS}), we obtain for}%
\begin{equation}
\bar{g}_{s,\chi}=-es\tilde{\mathbf{E}}\cdot\left\langle \boldsymbol{\upsilon
}_{s,n}^{\chi}\right\rangle _{s,\chi}\frac{\tau_{\mathrm{inter}}^{\mathrm{c}%
}\tau_{\mathrm{inter}}^{\mathrm{s}}}{\tau_{\mathrm{inter}}^{\mathrm{c}}%
+\tau_{\mathrm{inter}}^{\mathrm{s}}}. \label{eq_duc}%
\end{equation}
As analyzed above, $\bar{g}_{s,\chi}$ now is only $\chi$-dependent and the
chemical potential difference $\Delta\mu_{z}=(\bar{g}_{s,+}-\bar{g}_{s,-})/2$
becomes spin-independent. Therefore, upon application of the spin bias, the
fermion population decreases in the left DP and increases in the right, as
illustrated in Fig. \ref{figLLs}(d). The overall effect of this process is
that the Dirac fermions are pumped from one DP to the other, which exhibits
the $\mathbb{Z}_{2}$ anomaly. Consequently, we dub $\Delta\mu_{z}$ the
$\mathbb{Z}_{2}$ chemical potential. A nonzero $\Delta\mu_{z}$ presented in
Figs. \ref{figLLs}(d)-(f) indicates that an imbalance of carrier density is
established between the two DPs. Usually, the spin-flip inter-Dirac-valley
relaxation is much slower than the other relaxation processes and thus, the
$\mathbb{Z}_{2}$ chemical potential is insensitive to the local magnetic disorder.

\emph{Positive LMC and LSMC- }The spin-dependent current density is given by
\begin{equation}
\boldsymbol{j}_{s}=\frac{e}{2\pi}%
{\displaystyle\sum\limits_{\chi}}
{\displaystyle\sum\limits_{n,k_{y}}}
\int g_{s,n}^{\chi}(\mathbf{k})\boldsymbol{\upsilon}_{s,n}^{\chi}\left[
\frac{\partial f_{0}(\varepsilon_{s,n}^{\chi})}{\partial\varepsilon
_{s,n}^{\chi}}\right]  dk_{z}\label{eq_cds}%
\end{equation}
with $g_{s,n}^{\chi}(\mathbf{k})=\mathbf{\nabla}\mu_{s}\cdot
\boldsymbol{\upsilon}_{s,n}^{\chi}\tau_{\mathrm{intra}}+\bar{g}_{s,\chi}$.
Incorporating the chiral and $\mathbb{Z}_{2}$ anomalies, {together with }%
$\sum_{s\chi}\bar{g}_{s,\chi}=0$ due to particle conservation of the system,
we average both sides of Eq. {(\ref{eq_gfun}) at the Fermi level }and obtain
\begin{equation}
\bar{g}_{s,\chi}=-s\chi\tilde{P}_{-}\frac{\tau_{\mathrm{inter}}^{\mathrm{c}%
}\tau_{\mathrm{intra}}^{\mathrm{s}}}{\tau_{\mathrm{inter}}^{\mathrm{c}}%
+\tau_{\mathrm{intra}}^{\mathrm{s}}}+\chi\tilde{P}_{+}\frac{\tau
_{\mathrm{inter}}^{\mathrm{c}}\tau_{\mathrm{inter}}^{\mathrm{s}}}%
{\tau_{\mathrm{inter}}^{\mathrm{c}}+\tau_{\mathrm{inter}}^{\mathrm{s}}%
},\label{eq_gsx}%
\end{equation}
where%
\begin{equation}
\tilde{P}_{\pm}=\frac{\mathbf{\nabla}\mu_{-}\cdot\left\langle
\boldsymbol{\upsilon}_{-,n}^{+}\right\rangle _{-,+}\pm\mathbf{\nabla}\mu
_{+}\cdot\left\langle \boldsymbol{\upsilon}_{+,n}^{+}\right\rangle _{+,+}}{2}.
\end{equation}
At low temperatures, one can further derive $\left\langle \boldsymbol{\upsilon
}_{s,n}^{\chi}\right\rangle _{s,\chi}=s\chi\tilde{\upsilon}_{\mathrm{F}%
}/\varTheta_{s}$, with $\varTheta_{s}=2{\sum\limits_{n=0}^{n_{c}}}\frac{{1}%
}{\lambda_{n,s}}-1$, where $\lambda_{n,s}=\sqrt{1-2|n|(\frac{\hbar\omega_{c}%
}{E_{\mathrm{F}}^{s}})^{2}}$ and $n_{c}=\mbox{sgn}(E_{\mathrm{F}%
})\mbox{int}\left[  (E_{\mathrm{F}}^{s})^{2}/2(\hbar\omega_{c})^{2}\right]  $
is level index of the highest (lowest) LL crossed by the Fermi level for
$E_{\mathrm{F}}>0$ ($E_{\mathrm{F}}<0$). To see the physical meaning of
$\tilde{P}_{\pm}$ more clearly, we set $m_{2}=0$ and then $\varTheta_{s}$ is
spin-independent. For $\mathbf{\nabla}\mu_{s}=-e\mathbf{E}${, }$\tilde{P}%
_{+}=0$ and Eq. {(\ref{eq_gsx}) returns to }Eq. {(\ref{eq_dus}), while for
}$\mathbf{\nabla}\mu_{s}=-es\tilde{\mathbf{E}}$, $\tilde{P}_{-}=0$ and Eq.
{(\ref{eq_gsx}) }recovers Eq. {(\ref{eq_duc}). Therefore, }$|\tilde{P}_{\pm}|$
in fact describes the effective power of the particle pumping between the DPs.

Substituting Eq. {(\ref{eq_gsx}) into }Eq. {(\ref{eq_cds}), we can express the
spin-dependent current density as}%
\begin{align}
j_{s,z}(B)  &  =\sigma_{D}^{s}(E_{z}+s\tilde{E}_{z})+\Delta\sigma
(B)\nonumber\\
&  \times\left(  \frac{\tau_{\mathrm{intra}}^{\mathrm{s}}}{\tau
_{\mathrm{inter}}^{\mathrm{c}}+\tau_{\mathrm{intra}}^{\mathrm{s}}}E_{z}%
+\frac{\tau_{\mathrm{inter}}^{\mathrm{s}}}{\tau_{\mathrm{inter}}^{\mathrm{c}%
}+\tau_{\mathrm{inter}}^{\mathrm{s}}}s\tilde{E}_{z}\right)  , \label{eq_jsz}%
\end{align}
where $\sigma_{D}^{s}=\frac{e^{2}n_{e}^{s}}{\hbar k_{\mathrm{F},s}}%
\tilde{\upsilon}_{\mathrm{F}}\tau_{\mathrm{intra}}$ is the Drude conductivity
with $k_{\mathrm{F},s}=|E_{\mathrm{F}}^{s}|/\hbar\tilde{\upsilon}_{\mathrm{F}%
}$ and $n_{e}^{s}=(1/3\pi^{2})k_{\mathrm{F},s}^{3}$ as the spin-resolved
carrier density, and%
\begin{equation}
\Delta\sigma(B)=\frac{e^{2}}{h}\frac{eB\tilde{\upsilon}_{\mathrm{F}}%
\tau_{\mathrm{inter}}^{\mathrm{c}}}{h}\sum_{s}\frac{1}{\varTheta_{s}}
\label{eq_dsB}%
\end{equation}
is the magnetoconductivity attributable to the nonequilibium local chemical
potentials. Equations {(\ref{eq_jsz}) is }the central result of our work, from
which{ we define the }spin-resolved {electric and spin }conductivity as
$\sigma_{c}(B)=j_{s,z}(B)/E_{z}$ and $s(B)=j_{s,z}(B)/s\tilde{E}_{z}$. The LMC
for Weyl semimetals of a single pair of Weyl nodes is given by Eq.
(\ref{eq_dsB}), which has been discussed by us in Ref.
\cite{PhysRevLett.122.036601}. Here, the electron orbital angular momentum
couples strongly with the magnetic field, and the Weyl fermions can relax via
electron scattering between multiple Fermi pockets. Therefore, new
characteristic will emerge in the magnetotransport. The LSMC, given by $\Delta s(B)\equiv\lbrack
s(B)-s(0)]$, {is a spin version of the LMC due to the }$\mathbb{Z}_{2}$
anomaly.

In Fig. \ref{figcon}, we plot the calculated $\Delta\sigma_{c}(B)\equiv
\lbrack\sigma_{c}(B)-\sigma_{c}(0)]$, $\Delta s(B)$ and $\Delta\sigma_{D}%
=\sum_{s}\sigma_{D}^{s}-\sigma_{D}$ as functions of $B$, where $\sigma_{D}$ is
the zero-field Drude conductivity. As seen from Figs. \ref{figcon}(a) and (b),
though the Zeeman effect is neglected, the spin degeneracy of the LMC and LSMC
are eliminated by the coupling of the electron orbital angular momentum and
magnetic field. Due to the chiral and $\mathbb{Z}_{2}$ anomalies, the LMC and
LSMC exhibit synchronous oscillations with the chiral and $\mathbb{Z}_{2}$
chemical potentials. The envelopes of the {oscillations are scaled with
}$B^{2}$ for $\hbar\omega_{c}\ll|E_{\mathrm{F}}|$, which is consistent with
the classical formula obtained in Refs.
\cite{PhysRevB.88.104412,PhysRevB.96.165101}. From Fig. \ref{figcon}(c), we
see that, because of the coupling between the electron orbital angular
momentum and magnetic field, the trivial Drude conductivity also contributes a
$B$-dependent term%
\begin{equation}
\Delta\sigma_{D}^{s}/\sigma_{D}=(\frac{em_{2}}{\hbar E_{\mathrm{F}}})^{2}%
B^{2}+2s\frac{em_{2}}{\hbar E_{\mathrm{F}}}B\
\end{equation}
to the positive LMC, which is similar to that due to the chiral anomaly.
Therefore, the $B^{2}$ or $B$ dependence emerging in the positive LMC is not
unique to the chiral anomaly. However, the quantum oscillations of the LMC are
originated from the chiral $n=0$ LLs, manifesting the chiral anomaly. As shown
by Eq.\ {(\ref{eq_jsz}), }the LSMC possesses all characters of the
chiral-anomaly-induced LMC, including the periodic-in-$1/B$ quantum
oscillations, as exhibited in the inset of Fig.\ \ref{figcon}(b). While the
chiral anomaly is very sensitive to the local magnetic impurities, please see
the inset of Fig. \ref{figcon}(a), the $\mathbb{Z}_{2}$ anomaly is immune to
local magnetic disorder.

In conclusion, we have theoretically studied the anomalous magnetotransports
in Dirac semimetals carrying the $\mathbb{Z}_{2}$ topological charge. We find
that a spin bias along the magnetic field can realize the $\mathbb{Z}_{2}$
anomaly for topological Dirac semimetals. Accompanied with this, there emerges
a measurable positive LSMC. We further demonstrate that the $B^{2}$ and $B$
dependences of the positive LMC are not unique to the chiral anomaly, because
similar field dependences can also originate from the coupling
between the electron orbital angular momentum and magnetic field. The
$\mathbb{Z}_{2}$ anomaly induced LSMC possesses all characters of the LMC due
to the chiral anomaly, and we argue that the periodic-in-$1/B$ quantum
oscillations superposed on the positive LSMC can serve as a fingerprint of the
$\mathbb{Z}_{2}$ anomaly in topological Dirac semimetals.

This work was supported by the National Natural Science Foundation of China
under Grants No. 11674160 (L.S.), 11874016 (R.-Q.W), 11804130 (W.L.) and 11574155 (R.M.), by the Key Program for Guangdong NSF of China under Grant No. 2017B030311003, GDUPS(2017) and the
project funded by South China Normal University under Grant No. 671215 and 8S0532.

\bibliographystyle{apsrev4-1}
\bibliography{bibNa3Bi}

\begin{thebibliography}{35}%
\makeatletter
\providecommand \@ifxundefined [1]{%
 \@ifx{#1\undefined}
}%
\providecommand \@ifnum [1]{%
 \ifnum #1\expandafter \@firstoftwo
 \else \expandafter \@secondoftwo
 \fi
}%
\providecommand \@ifx [1]{%
 \ifx #1\expandafter \@firstoftwo
 \else \expandafter \@secondoftwo
 \fi
}%
\providecommand \natexlab [1]{#1}%
\providecommand \enquote  [1]{``#1''}%
\providecommand \bibnamefont  [1]{#1}%
\providecommand \bibfnamefont [1]{#1}%
\providecommand \citenamefont [1]{#1}%
\providecommand \href@noop [0]{\@secondoftwo}%
\providecommand \href [0]{\begingroup \@sanitize@url \@href}%
\providecommand \@href[1]{\@@startlink{#1}\@@href}%
\providecommand \@@href[1]{\endgroup#1\@@endlink}%
\providecommand \@sanitize@url [0]{\catcode `\\12\catcode `\$12\catcode
  `\&12\catcode `\#12\catcode `\^12\catcode `\_12\catcode `\%12\relax}%
\providecommand \@@startlink[1]{}%
\providecommand \@@endlink[0]{}%
\providecommand \url  [0]{\begingroup\@sanitize@url \@url }%
\providecommand \@url [1]{\endgroup\@href {#1}{\urlprefix }}%
\providecommand \urlprefix  [0]{URL }%
\providecommand \Eprint [0]{\href }%
\providecommand \doibase [0]{http://dx.doi.org/}%
\providecommand \selectlanguage [0]{\@gobble}%
\providecommand \bibinfo  [0]{\@secondoftwo}%
\providecommand \bibfield  [0]{\@secondoftwo}%
\providecommand \translation [1]{[#1]}%
\providecommand \BibitemOpen [0]{}%
\providecommand \bibitemStop [0]{}%
\providecommand \bibitemNoStop [0]{.\EOS\space}%
\providecommand \EOS [0]{\spacefactor3000\relax}%
\providecommand \BibitemShut  [1]{\csname bibitem#1\endcsname}%
\let\auto@bib@innerbib\@empty
\bibitem [{\citenamefont {Armitage}\ \emph {et~al.}(2018)\citenamefont
  {Armitage}, \citenamefont {Mele},\ and\ \citenamefont
  {Vishwanath}}]{RevModPhys.90.015001}%
  \BibitemOpen
  \bibfield  {author} {\bibinfo {author} {\bibfnamefont {N.~P.}\ \bibnamefont
  {Armitage}}, \bibinfo {author} {\bibfnamefont {E.~J.}\ \bibnamefont {Mele}},
  \ and\ \bibinfo {author} {\bibfnamefont {A.}~\bibnamefont {Vishwanath}},\
  }\href {\doibase 10.1103/RevModPhys.90.015001} {\bibfield  {journal}
  {\bibinfo  {journal} {Rev. Mod. Phys.}\ }\textbf {\bibinfo {volume} {90}},\
  \bibinfo {pages} {015001} (\bibinfo {year} {2018})}\BibitemShut {NoStop}%
\bibitem [{\citenamefont {Liu}\ \emph {et~al.}(2014)\citenamefont {Liu},
  \citenamefont {Zhou}, \citenamefont {Zhang}, \citenamefont {Wang},
  \citenamefont {Weng}, \citenamefont {Prabhakaran}, \citenamefont {Mo},
  \citenamefont {Shen}, \citenamefont {Fang}, \citenamefont {Dai},
  \citenamefont {Hussain},\ and\ \citenamefont {Chen}}]{Liu864}%
  \BibitemOpen
  \bibfield  {author} {\bibinfo {author} {\bibfnamefont {Z.~K.}\ \bibnamefont
  {Liu}}, \bibinfo {author} {\bibfnamefont {B.}~\bibnamefont {Zhou}}, \bibinfo
  {author} {\bibfnamefont {Y.}~\bibnamefont {Zhang}}, \bibinfo {author}
  {\bibfnamefont {Z.~J.}\ \bibnamefont {Wang}}, \bibinfo {author}
  {\bibfnamefont {H.~M.}\ \bibnamefont {Weng}}, \bibinfo {author}
  {\bibfnamefont {D.}~\bibnamefont {Prabhakaran}}, \bibinfo {author}
  {\bibfnamefont {S.-K.}\ \bibnamefont {Mo}}, \bibinfo {author} {\bibfnamefont
  {Z.~X.}\ \bibnamefont {Shen}}, \bibinfo {author} {\bibfnamefont
  {Z.}~\bibnamefont {Fang}}, \bibinfo {author} {\bibfnamefont {X.}~\bibnamefont
  {Dai}}, \bibinfo {author} {\bibfnamefont {Z.}~\bibnamefont {Hussain}}, \ and\
  \bibinfo {author} {\bibfnamefont {Y.~L.}\ \bibnamefont {Chen}},\ }\href
  {\doibase 10.1126/science.1245085} {\bibfield  {journal} {\bibinfo  {journal}
  {Science}\ }\textbf {\bibinfo {volume} {343}},\ \bibinfo {pages} {864}
  (\bibinfo {year} {2014})}\BibitemShut {NoStop}%
\bibitem [{\citenamefont {Zhang}\ \emph {et~al.}(2016)\citenamefont {Zhang},
  \citenamefont {Xu}, \citenamefont {Belopolski}, \citenamefont {Yuan},
  \citenamefont {Lin}, \citenamefont {Tong}, \citenamefont {Bian},
  \citenamefont {Alidoust}, \citenamefont {Lee}, \citenamefont {Huang},
  \citenamefont {Chang}, \citenamefont {Chang}, \citenamefont {Hsu},
  \citenamefont {Jeng}, \citenamefont {Neupane}, \citenamefont {Sanchez},
  \citenamefont {Zheng}, \citenamefont {Wang}, \citenamefont {Lin},
  \citenamefont {Zhang}, \citenamefont {Lu}, \citenamefont {Shen},
  \citenamefont {Neupert}, \citenamefont {Zahid~Hasan},\ and\ \citenamefont
  {Jia}}]{Zhang:2016aa}%
  \BibitemOpen
  \bibfield  {author} {\bibinfo {author} {\bibfnamefont {C.-L.}\ \bibnamefont
  {Zhang}}, \bibinfo {author} {\bibfnamefont {S.-Y.}\ \bibnamefont {Xu}},
  \bibinfo {author} {\bibfnamefont {I.}~\bibnamefont {Belopolski}}, \bibinfo
  {author} {\bibfnamefont {Z.}~\bibnamefont {Yuan}}, \bibinfo {author}
  {\bibfnamefont {Z.}~\bibnamefont {Lin}}, \bibinfo {author} {\bibfnamefont
  {B.}~\bibnamefont {Tong}}, \bibinfo {author} {\bibfnamefont {G.}~\bibnamefont
  {Bian}}, \bibinfo {author} {\bibfnamefont {N.}~\bibnamefont {Alidoust}},
  \bibinfo {author} {\bibfnamefont {C.-C.}\ \bibnamefont {Lee}}, \bibinfo
  {author} {\bibfnamefont {S.-M.}\ \bibnamefont {Huang}}, \bibinfo {author}
  {\bibfnamefont {T.-R.}\ \bibnamefont {Chang}}, \bibinfo {author}
  {\bibfnamefont {G.}~\bibnamefont {Chang}}, \bibinfo {author} {\bibfnamefont
  {C.-H.}\ \bibnamefont {Hsu}}, \bibinfo {author} {\bibfnamefont {H.-T.}\
  \bibnamefont {Jeng}}, \bibinfo {author} {\bibfnamefont {M.}~\bibnamefont
  {Neupane}}, \bibinfo {author} {\bibfnamefont {D.~S.}\ \bibnamefont
  {Sanchez}}, \bibinfo {author} {\bibfnamefont {H.}~\bibnamefont {Zheng}},
  \bibinfo {author} {\bibfnamefont {J.}~\bibnamefont {Wang}}, \bibinfo {author}
  {\bibfnamefont {H.}~\bibnamefont {Lin}}, \bibinfo {author} {\bibfnamefont
  {C.}~\bibnamefont {Zhang}}, \bibinfo {author} {\bibfnamefont {H.-Z.}\
  \bibnamefont {Lu}}, \bibinfo {author} {\bibfnamefont {S.-Q.}\ \bibnamefont
  {Shen}}, \bibinfo {author} {\bibfnamefont {T.}~\bibnamefont {Neupert}},
  \bibinfo {author} {\bibfnamefont {M.}~\bibnamefont {Zahid~Hasan}}, \ and\
  \bibinfo {author} {\bibfnamefont {S.}~\bibnamefont {Jia}},\ }\href
  {https://doi.org/10.1038/ncomms10735} {\bibfield  {journal} {\bibinfo
  {journal} {Nat. Commun.}\ }\textbf {\bibinfo {volume} {7}},\ \bibinfo {pages}
  {10735} (\bibinfo {year} {2016})}\BibitemShut {NoStop}%
\bibitem [{\citenamefont {Xiong}\ \emph {et~al.}(2015)\citenamefont {Xiong},
  \citenamefont {Kushwaha}, \citenamefont {Liang}, \citenamefont {Krizan},
  \citenamefont {Hirschberger}, \citenamefont {Wang}, \citenamefont {Cava},\
  and\ \citenamefont {Ong}}]{Xiong413}%
  \BibitemOpen
  \bibfield  {author} {\bibinfo {author} {\bibfnamefont {J.}~\bibnamefont
  {Xiong}}, \bibinfo {author} {\bibfnamefont {S.~K.}\ \bibnamefont {Kushwaha}},
  \bibinfo {author} {\bibfnamefont {T.}~\bibnamefont {Liang}}, \bibinfo
  {author} {\bibfnamefont {J.~W.}\ \bibnamefont {Krizan}}, \bibinfo {author}
  {\bibfnamefont {M.}~\bibnamefont {Hirschberger}}, \bibinfo {author}
  {\bibfnamefont {W.}~\bibnamefont {Wang}}, \bibinfo {author} {\bibfnamefont
  {R.~J.}\ \bibnamefont {Cava}}, \ and\ \bibinfo {author} {\bibfnamefont
  {N.~P.}\ \bibnamefont {Ong}},\ }\href {\doibase 10.1126/science.aac6089}
  {\bibfield  {journal} {\bibinfo  {journal} {Science}\ }\textbf {\bibinfo
  {volume} {350}},\ \bibinfo {pages} {413} (\bibinfo {year}
  {2015})}\BibitemShut {NoStop}%
\bibitem [{\citenamefont {Zhang}\ \emph {et~al.}(2017)\citenamefont {Zhang},
  \citenamefont {Zhang}, \citenamefont {Wang}, \citenamefont {Liu},
  \citenamefont {Chen}, \citenamefont {Lu}, \citenamefont {Liang},
  \citenamefont {Cao}, \citenamefont {Yuan}, \citenamefont {Tang},
  \citenamefont {Li}, \citenamefont {Zhou}, \citenamefont {Gu}, \citenamefont
  {Wu}, \citenamefont {Zou},\ and\ \citenamefont {Xiu}}]{Zhang:2017aa}%
  \BibitemOpen
  \bibfield  {author} {\bibinfo {author} {\bibfnamefont {C.}~\bibnamefont
  {Zhang}}, \bibinfo {author} {\bibfnamefont {E.}~\bibnamefont {Zhang}},
  \bibinfo {author} {\bibfnamefont {W.}~\bibnamefont {Wang}}, \bibinfo {author}
  {\bibfnamefont {Y.}~\bibnamefont {Liu}}, \bibinfo {author} {\bibfnamefont
  {Z.-G.}\ \bibnamefont {Chen}}, \bibinfo {author} {\bibfnamefont
  {S.}~\bibnamefont {Lu}}, \bibinfo {author} {\bibfnamefont {S.}~\bibnamefont
  {Liang}}, \bibinfo {author} {\bibfnamefont {J.}~\bibnamefont {Cao}}, \bibinfo
  {author} {\bibfnamefont {X.}~\bibnamefont {Yuan}}, \bibinfo {author}
  {\bibfnamefont {L.}~\bibnamefont {Tang}}, \bibinfo {author} {\bibfnamefont
  {Q.}~\bibnamefont {Li}}, \bibinfo {author} {\bibfnamefont {C.}~\bibnamefont
  {Zhou}}, \bibinfo {author} {\bibfnamefont {T.}~\bibnamefont {Gu}}, \bibinfo
  {author} {\bibfnamefont {Y.}~\bibnamefont {Wu}}, \bibinfo {author}
  {\bibfnamefont {J.}~\bibnamefont {Zou}}, \ and\ \bibinfo {author}
  {\bibfnamefont {F.}~\bibnamefont {Xiu}},\ }\href
  {https://doi.org/10.1038/ncomms13741} {\bibfield  {journal} {\bibinfo
  {journal} {Nat. Commun.}\ }\textbf {\bibinfo {volume} {8}},\ \bibinfo {pages}
  {13741} (\bibinfo {year} {2017})}\BibitemShut {NoStop}%
\bibitem [{\citenamefont {Wang}\ \emph {et~al.}(2017)\citenamefont {Wang},
  \citenamefont {Sun}, \citenamefont {Lu},\ and\ \citenamefont
  {Xie}}]{PhysRevLett.119.136806}%
  \BibitemOpen
  \bibfield  {author} {\bibinfo {author} {\bibfnamefont {C.~M.}\ \bibnamefont
  {Wang}}, \bibinfo {author} {\bibfnamefont {H.-P.}\ \bibnamefont {Sun}},
  \bibinfo {author} {\bibfnamefont {H.-Z.}\ \bibnamefont {Lu}}, \ and\ \bibinfo
  {author} {\bibfnamefont {X.~C.}\ \bibnamefont {Xie}},\ }\href {\doibase
  10.1103/PhysRevLett.119.136806} {\bibfield  {journal} {\bibinfo  {journal}
  {Phys. Rev. Lett.}\ }\textbf {\bibinfo {volume} {119}},\ \bibinfo {pages}
  {136806} (\bibinfo {year} {2017})}\BibitemShut {NoStop}%
\bibitem [{\citenamefont {Yang}\ and\ \citenamefont
  {Nagaosa}(2014)}]{Yang:2014aa}%
  \BibitemOpen
  \bibfield  {author} {\bibinfo {author} {\bibfnamefont {B.-J.}\ \bibnamefont
  {Yang}}\ and\ \bibinfo {author} {\bibfnamefont {N.}~\bibnamefont {Nagaosa}},\
  }\href {https://doi.org/10.1038/ncomms5898} {\bibfield  {journal} {\bibinfo
  {journal} {Nat. Commun.}\ }\textbf {\bibinfo {volume} {5}},\ \bibinfo {pages}
  {4898} (\bibinfo {year} {2014})}\BibitemShut {NoStop}%
\bibitem [{\citenamefont {Gorbar}\ \emph {et~al.}(2015)\citenamefont {Gorbar},
  \citenamefont {Miransky}, \citenamefont {Shovkovy},\ and\ \citenamefont
  {Sukhachov}}]{PhysRevB.91.121101}%
  \BibitemOpen
  \bibfield  {author} {\bibinfo {author} {\bibfnamefont {E.~V.}\ \bibnamefont
  {Gorbar}}, \bibinfo {author} {\bibfnamefont {V.~A.}\ \bibnamefont
  {Miransky}}, \bibinfo {author} {\bibfnamefont {I.~A.}\ \bibnamefont
  {Shovkovy}}, \ and\ \bibinfo {author} {\bibfnamefont {P.~O.}\ \bibnamefont
  {Sukhachov}},\ }\href {\doibase 10.1103/PhysRevB.91.121101} {\bibfield
  {journal} {\bibinfo  {journal} {Phys. Rev. B}\ }\textbf {\bibinfo {volume}
  {91}},\ \bibinfo {pages} {121101} (\bibinfo {year} {2015})}\BibitemShut
  {NoStop}%
\bibitem [{\citenamefont {Kargarian}\ \emph {et~al.}(2016)\citenamefont
  {Kargarian}, \citenamefont {Randeria},\ and\ \citenamefont
  {Lu}}]{Kargarian8648}%
  \BibitemOpen
  \bibfield  {author} {\bibinfo {author} {\bibfnamefont {M.}~\bibnamefont
  {Kargarian}}, \bibinfo {author} {\bibfnamefont {M.}~\bibnamefont {Randeria}},
  \ and\ \bibinfo {author} {\bibfnamefont {Y.-M.}\ \bibnamefont {Lu}},\ }\href
  {\doibase 10.1073/pnas.1524787113} {\bibfield  {journal} {\bibinfo  {journal}
  {Proceedings of the National Academy of Sciences}\ }\textbf {\bibinfo
  {volume} {113}},\ \bibinfo {pages} {8648} (\bibinfo {year}
  {2016})}\BibitemShut {NoStop}%
\bibitem [{\citenamefont {Raza}\ \emph {et~al.}(2019)\citenamefont {Raza},
  \citenamefont {Sirota},\ and\ \citenamefont {Teo}}]{PhysRevX.9.011039}%
  \BibitemOpen
  \bibfield  {author} {\bibinfo {author} {\bibfnamefont {S.}~\bibnamefont
  {Raza}}, \bibinfo {author} {\bibfnamefont {A.}~\bibnamefont {Sirota}}, \ and\
  \bibinfo {author} {\bibfnamefont {J.~C.~Y.}\ \bibnamefont {Teo}},\ }\href
  {\doibase 10.1103/PhysRevX.9.011039} {\bibfield  {journal} {\bibinfo
  {journal} {Phys. Rev. X}\ }\textbf {\bibinfo {volume} {9}},\ \bibinfo {pages}
  {011039} (\bibinfo {year} {2019})}\BibitemShut {NoStop}%
\bibitem [{\citenamefont {Zyuzin}\ \emph {et~al.}(2012)\citenamefont {Zyuzin},
  \citenamefont {Wu},\ and\ \citenamefont {Burkov}}]{PhysRevB.85.165110}%
  \BibitemOpen
  \bibfield  {author} {\bibinfo {author} {\bibfnamefont {A.~A.}\ \bibnamefont
  {Zyuzin}}, \bibinfo {author} {\bibfnamefont {S.}~\bibnamefont {Wu}}, \ and\
  \bibinfo {author} {\bibfnamefont {A.~A.}\ \bibnamefont {Burkov}},\ }\href
  {\doibase 10.1103/PhysRevB.85.165110} {\bibfield  {journal} {\bibinfo
  {journal} {Phys. Rev. B}\ }\textbf {\bibinfo {volume} {85}},\ \bibinfo
  {pages} {165110} (\bibinfo {year} {2012})}\BibitemShut {NoStop}%
\bibitem [{\citenamefont {Goswami}\ and\ \citenamefont
  {Tewari}(2013)}]{PhysRevB.88.245107}%
  \BibitemOpen
  \bibfield  {author} {\bibinfo {author} {\bibfnamefont {P.}~\bibnamefont
  {Goswami}}\ and\ \bibinfo {author} {\bibfnamefont {S.}~\bibnamefont
  {Tewari}},\ }\href {\doibase 10.1103/PhysRevB.88.245107} {\bibfield
  {journal} {\bibinfo  {journal} {Phys. Rev. B}\ }\textbf {\bibinfo {volume}
  {88}},\ \bibinfo {pages} {245107} (\bibinfo {year} {2013})}\BibitemShut
  {NoStop}%
\bibitem [{\citenamefont {Han}\ \emph {et~al.}(2018)\citenamefont {Han},
  \citenamefont {Cho},\ and\ \citenamefont {Moon}}]{PhysRevB.98.085149}%
  \BibitemOpen
  \bibfield  {author} {\bibinfo {author} {\bibfnamefont {S.}~\bibnamefont
  {Han}}, \bibinfo {author} {\bibfnamefont {G.~Y.}\ \bibnamefont {Cho}}, \ and\
  \bibinfo {author} {\bibfnamefont {E.-G.}\ \bibnamefont {Moon}},\ }\href
  {\doibase 10.1103/PhysRevB.98.085149} {\bibfield  {journal} {\bibinfo
  {journal} {Phys. Rev. B}\ }\textbf {\bibinfo {volume} {98}},\ \bibinfo
  {pages} {085149} (\bibinfo {year} {2018})}\BibitemShut {NoStop}%
\bibitem [{\citenamefont {Deng}\ \emph {et~al.}(2017)\citenamefont {Deng},
  \citenamefont {Luo}, \citenamefont {Wang}, \citenamefont {Sheng},\ and\
  \citenamefont {Xing}}]{PhysRevB.96.155141}%
  \BibitemOpen
  \bibfield  {author} {\bibinfo {author} {\bibfnamefont {M.-X.}\ \bibnamefont
  {Deng}}, \bibinfo {author} {\bibfnamefont {W.}~\bibnamefont {Luo}}, \bibinfo
  {author} {\bibfnamefont {R.-Q.}\ \bibnamefont {Wang}}, \bibinfo {author}
  {\bibfnamefont {L.}~\bibnamefont {Sheng}}, \ and\ \bibinfo {author}
  {\bibfnamefont {D.~Y.}\ \bibnamefont {Xing}},\ }\href {\doibase
  10.1103/PhysRevB.96.155141} {\bibfield  {journal} {\bibinfo  {journal} {Phys.
  Rev. B}\ }\textbf {\bibinfo {volume} {96}},\ \bibinfo {pages} {155141}
  (\bibinfo {year} {2017})}\BibitemShut {NoStop}%
\bibitem [{\citenamefont {Chen}\ \emph {et~al.}(2019)\citenamefont {Chen},
  \citenamefont {Yu}, \citenamefont {Li}, \citenamefont {Chen}, \citenamefont
  {Sheng},\ and\ \citenamefont {Yang}}]{PhysRevB.99.075131}%
  \BibitemOpen
  \bibfield  {author} {\bibinfo {author} {\bibfnamefont {C.}~\bibnamefont
  {Chen}}, \bibinfo {author} {\bibfnamefont {Z.-M.}\ \bibnamefont {Yu}},
  \bibinfo {author} {\bibfnamefont {S.}~\bibnamefont {Li}}, \bibinfo {author}
  {\bibfnamefont {Z.}~\bibnamefont {Chen}}, \bibinfo {author} {\bibfnamefont
  {X.-L.}\ \bibnamefont {Sheng}}, \ and\ \bibinfo {author} {\bibfnamefont
  {S.~A.}\ \bibnamefont {Yang}},\ }\href {\doibase 10.1103/PhysRevB.99.075131}
  {\bibfield  {journal} {\bibinfo  {journal} {Phys. Rev. B}\ }\textbf {\bibinfo
  {volume} {99}},\ \bibinfo {pages} {075131} (\bibinfo {year}
  {2019})}\BibitemShut {NoStop}%
\bibitem [{\citenamefont {Nielsen}\ and\ \citenamefont
  {Ninomiya}(1983)}]{NIELSEN1983389}%
  \BibitemOpen
  \bibfield  {author} {\bibinfo {author} {\bibfnamefont {H.}~\bibnamefont
  {Nielsen}}\ and\ \bibinfo {author} {\bibfnamefont {M.}~\bibnamefont
  {Ninomiya}},\ }\href {\doibase https://doi.org/10.1016/0370-2693(83)91529-0}
  {\bibfield  {journal} {\bibinfo  {journal} {Physics Letters B}\ }\textbf
  {\bibinfo {volume} {130}},\ \bibinfo {pages} {389 } (\bibinfo {year}
  {1983})}\BibitemShut {NoStop}%
\bibitem [{\citenamefont {Volovik}(2003)}]{Volovik2003}%
  \BibitemOpen
  \bibfield  {author} {\bibinfo {author} {\bibfnamefont {G.~E.}\ \bibnamefont
  {Volovik}},\ }\href@noop {} {\emph {\bibinfo {title} {The universe in a
  helium droplet}}},\ Vol.\ \bibinfo {volume} {117}\ (\bibinfo  {publisher}
  {Oxford University Press on Demand},\ \bibinfo {year} {2003})\BibitemShut
  {NoStop}%
\bibitem [{\citenamefont {Wan}\ \emph {et~al.}(2011)\citenamefont {Wan},
  \citenamefont {Turner}, \citenamefont {Vishwanath},\ and\ \citenamefont
  {Savrasov}}]{PhysRevB.83.205101}%
  \BibitemOpen
  \bibfield  {author} {\bibinfo {author} {\bibfnamefont {X.}~\bibnamefont
  {Wan}}, \bibinfo {author} {\bibfnamefont {A.~M.}\ \bibnamefont {Turner}},
  \bibinfo {author} {\bibfnamefont {A.}~\bibnamefont {Vishwanath}}, \ and\
  \bibinfo {author} {\bibfnamefont {S.~Y.}\ \bibnamefont {Savrasov}},\ }\href
  {\doibase 10.1103/PhysRevB.83.205101} {\bibfield  {journal} {\bibinfo
  {journal} {Phys. Rev. B}\ }\textbf {\bibinfo {volume} {83}},\ \bibinfo
  {pages} {205101} (\bibinfo {year} {2011})}\BibitemShut {NoStop}%
\bibitem [{\citenamefont {Huang}\ \emph {et~al.}(2015)\citenamefont {Huang},
  \citenamefont {Zhao}, \citenamefont {Long}, \citenamefont {Wang},
  \citenamefont {Chen}, \citenamefont {Yang}, \citenamefont {Liang},
  \citenamefont {Xue}, \citenamefont {Weng}, \citenamefont {Fang},
  \citenamefont {Dai},\ and\ \citenamefont {Chen}}]{PhysRevX.5.031023}%
  \BibitemOpen
  \bibfield  {author} {\bibinfo {author} {\bibfnamefont {X.}~\bibnamefont
  {Huang}}, \bibinfo {author} {\bibfnamefont {L.}~\bibnamefont {Zhao}},
  \bibinfo {author} {\bibfnamefont {Y.}~\bibnamefont {Long}}, \bibinfo {author}
  {\bibfnamefont {P.}~\bibnamefont {Wang}}, \bibinfo {author} {\bibfnamefont
  {D.}~\bibnamefont {Chen}}, \bibinfo {author} {\bibfnamefont {Z.}~\bibnamefont
  {Yang}}, \bibinfo {author} {\bibfnamefont {H.}~\bibnamefont {Liang}},
  \bibinfo {author} {\bibfnamefont {M.}~\bibnamefont {Xue}}, \bibinfo {author}
  {\bibfnamefont {H.}~\bibnamefont {Weng}}, \bibinfo {author} {\bibfnamefont
  {Z.}~\bibnamefont {Fang}}, \bibinfo {author} {\bibfnamefont {X.}~\bibnamefont
  {Dai}}, \ and\ \bibinfo {author} {\bibfnamefont {G.}~\bibnamefont {Chen}},\
  }\href {\doibase 10.1103/PhysRevX.5.031023} {\bibfield  {journal} {\bibinfo
  {journal} {Phys. Rev. X}\ }\textbf {\bibinfo {volume} {5}},\ \bibinfo {pages}
  {031023} (\bibinfo {year} {2015})}\BibitemShut {NoStop}%
\bibitem [{\citenamefont {Deng}\ \emph
  {et~al.}(2019{\natexlab{a}})\citenamefont {Deng}, \citenamefont {Qi},
  \citenamefont {Ma}, \citenamefont {Shen}, \citenamefont {Wang}, \citenamefont
  {Sheng},\ and\ \citenamefont {Xing}}]{PhysRevLett.122.036601}%
  \BibitemOpen
  \bibfield  {author} {\bibinfo {author} {\bibfnamefont {M.-X.}\ \bibnamefont
  {Deng}}, \bibinfo {author} {\bibfnamefont {G.~Y.}\ \bibnamefont {Qi}},
  \bibinfo {author} {\bibfnamefont {R.}~\bibnamefont {Ma}}, \bibinfo {author}
  {\bibfnamefont {R.}~\bibnamefont {Shen}}, \bibinfo {author} {\bibfnamefont
  {R.-Q.}\ \bibnamefont {Wang}}, \bibinfo {author} {\bibfnamefont
  {L.}~\bibnamefont {Sheng}}, \ and\ \bibinfo {author} {\bibfnamefont {D.~Y.}\
  \bibnamefont {Xing}},\ }\href {\doibase 10.1103/PhysRevLett.122.036601}
  {\bibfield  {journal} {\bibinfo  {journal} {Phys. Rev. Lett.}\ }\textbf
  {\bibinfo {volume} {122}},\ \bibinfo {pages} {036601} (\bibinfo {year}
  {2019}{\natexlab{a}})}\BibitemShut {NoStop}%
\bibitem [{\citenamefont {Liang}\ \emph {et~al.}(2018)\citenamefont {Liang},
  \citenamefont {Lin}, \citenamefont {Kushwaha}, \citenamefont {Xing},
  \citenamefont {Ni}, \citenamefont {Cava},\ and\ \citenamefont
  {Ong}}]{PhysRevX.8.031002}%
  \BibitemOpen
  \bibfield  {author} {\bibinfo {author} {\bibfnamefont {S.}~\bibnamefont
  {Liang}}, \bibinfo {author} {\bibfnamefont {J.}~\bibnamefont {Lin}}, \bibinfo
  {author} {\bibfnamefont {S.}~\bibnamefont {Kushwaha}}, \bibinfo {author}
  {\bibfnamefont {J.}~\bibnamefont {Xing}}, \bibinfo {author} {\bibfnamefont
  {N.}~\bibnamefont {Ni}}, \bibinfo {author} {\bibfnamefont {R.~J.}\
  \bibnamefont {Cava}}, \ and\ \bibinfo {author} {\bibfnamefont {N.~P.}\
  \bibnamefont {Ong}},\ }\href {\doibase 10.1103/PhysRevX.8.031002} {\bibfield
  {journal} {\bibinfo  {journal} {Phys. Rev. X}\ }\textbf {\bibinfo {volume}
  {8}},\ \bibinfo {pages} {031002} (\bibinfo {year} {2018})}\BibitemShut
  {NoStop}%
\bibitem [{\citenamefont {Neupane}\ \emph {et~al.}(2014)\citenamefont
  {Neupane}, \citenamefont {Xu}, \citenamefont {Sankar}, \citenamefont
  {Alidoust}, \citenamefont {Bian}, \citenamefont {Liu}, \citenamefont
  {Belopolski}, \citenamefont {Chang}, \citenamefont {Jeng}, \citenamefont
  {Lin}, \citenamefont {Bansil}, \citenamefont {Chou},\ and\ \citenamefont
  {Hasan}}]{Neupane:2014aa}%
  \BibitemOpen
  \bibfield  {author} {\bibinfo {author} {\bibfnamefont {M.}~\bibnamefont
  {Neupane}}, \bibinfo {author} {\bibfnamefont {S.-Y.}\ \bibnamefont {Xu}},
  \bibinfo {author} {\bibfnamefont {R.}~\bibnamefont {Sankar}}, \bibinfo
  {author} {\bibfnamefont {N.}~\bibnamefont {Alidoust}}, \bibinfo {author}
  {\bibfnamefont {G.}~\bibnamefont {Bian}}, \bibinfo {author} {\bibfnamefont
  {C.}~\bibnamefont {Liu}}, \bibinfo {author} {\bibfnamefont {I.}~\bibnamefont
  {Belopolski}}, \bibinfo {author} {\bibfnamefont {T.-R.}\ \bibnamefont
  {Chang}}, \bibinfo {author} {\bibfnamefont {H.-T.}\ \bibnamefont {Jeng}},
  \bibinfo {author} {\bibfnamefont {H.}~\bibnamefont {Lin}}, \bibinfo {author}
  {\bibfnamefont {A.}~\bibnamefont {Bansil}}, \bibinfo {author} {\bibfnamefont
  {F.}~\bibnamefont {Chou}}, \ and\ \bibinfo {author} {\bibfnamefont {M.~Z.}\
  \bibnamefont {Hasan}},\ }\href {https://doi.org/10.1038/ncomms4786}
  {\bibfield  {journal} {\bibinfo  {journal} {Nat. Commun.}\ }\textbf {\bibinfo
  {volume} {5}},\ \bibinfo {pages} {3786} (\bibinfo {year} {2014})}\BibitemShut
  {NoStop}%
\bibitem [{\citenamefont {Li}\ \emph {et~al.}(2015)\citenamefont {Li},
  \citenamefont {Wang}, \citenamefont {Liu}, \citenamefont {Wang},
  \citenamefont {Liao},\ and\ \citenamefont {Yu}}]{Li:2015aa}%
  \BibitemOpen
  \bibfield  {author} {\bibinfo {author} {\bibfnamefont {C.-Z.}\ \bibnamefont
  {Li}}, \bibinfo {author} {\bibfnamefont {L.-X.}\ \bibnamefont {Wang}},
  \bibinfo {author} {\bibfnamefont {H.}~\bibnamefont {Liu}}, \bibinfo {author}
  {\bibfnamefont {J.}~\bibnamefont {Wang}}, \bibinfo {author} {\bibfnamefont
  {Z.-M.}\ \bibnamefont {Liao}}, \ and\ \bibinfo {author} {\bibfnamefont
  {D.-P.}\ \bibnamefont {Yu}},\ }\href {https://doi.org/10.1038/ncomms10137}
  {\bibfield  {journal} {\bibinfo  {journal} {Nat. Commun.}\ }\textbf {\bibinfo
  {volume} {6}},\ \bibinfo {pages} {10137} (\bibinfo {year}
  {2015})}\BibitemShut {NoStop}%
\bibitem [{\citenamefont {Wang}\ \emph {et~al.}(2012)\citenamefont {Wang},
  \citenamefont {Sun}, \citenamefont {Chen}, \citenamefont {Franchini},
  \citenamefont {Xu}, \citenamefont {Weng}, \citenamefont {Dai},\ and\
  \citenamefont {Fang}}]{A3B_PhysRevB.85.195320}%
  \BibitemOpen
  \bibfield  {author} {\bibinfo {author} {\bibfnamefont {Z.}~\bibnamefont
  {Wang}}, \bibinfo {author} {\bibfnamefont {Y.}~\bibnamefont {Sun}}, \bibinfo
  {author} {\bibfnamefont {X.-Q.}\ \bibnamefont {Chen}}, \bibinfo {author}
  {\bibfnamefont {C.}~\bibnamefont {Franchini}}, \bibinfo {author}
  {\bibfnamefont {G.}~\bibnamefont {Xu}}, \bibinfo {author} {\bibfnamefont
  {H.}~\bibnamefont {Weng}}, \bibinfo {author} {\bibfnamefont {X.}~\bibnamefont
  {Dai}}, \ and\ \bibinfo {author} {\bibfnamefont {Z.}~\bibnamefont {Fang}},\
  }\href {\doibase 10.1103/PhysRevB.85.195320} {\bibfield  {journal} {\bibinfo
  {journal} {Phys. Rev. B}\ }\textbf {\bibinfo {volume} {85}},\ \bibinfo
  {pages} {195320} (\bibinfo {year} {2012})}\BibitemShut {NoStop}%
\bibitem [{\citenamefont {Wang}\ \emph {et~al.}(2013)\citenamefont {Wang},
  \citenamefont {Weng}, \citenamefont {Wu}, \citenamefont {Dai},\ and\
  \citenamefont {Fang}}]{Cd3As2_PhysRevB.88.125427}%
  \BibitemOpen
  \bibfield  {author} {\bibinfo {author} {\bibfnamefont {Z.}~\bibnamefont
  {Wang}}, \bibinfo {author} {\bibfnamefont {H.}~\bibnamefont {Weng}}, \bibinfo
  {author} {\bibfnamefont {Q.}~\bibnamefont {Wu}}, \bibinfo {author}
  {\bibfnamefont {X.}~\bibnamefont {Dai}}, \ and\ \bibinfo {author}
  {\bibfnamefont {Z.}~\bibnamefont {Fang}},\ }\href {\doibase
  10.1103/PhysRevB.88.125427} {\bibfield  {journal} {\bibinfo  {journal} {Phys.
  Rev. B}\ }\textbf {\bibinfo {volume} {88}},\ \bibinfo {pages} {125427}
  (\bibinfo {year} {2013})}\BibitemShut {NoStop}%
\bibitem [{\citenamefont {Burkov}\ and\ \citenamefont
  {Kim}(2016)}]{PhysRevLett.117.136602}%
  \BibitemOpen
  \bibfield  {author} {\bibinfo {author} {\bibfnamefont {A.~A.}\ \bibnamefont
  {Burkov}}\ and\ \bibinfo {author} {\bibfnamefont {Y.~B.}\ \bibnamefont
  {Kim}},\ }\href {\doibase 10.1103/PhysRevLett.117.136602} {\bibfield
  {journal} {\bibinfo  {journal} {Phys. Rev. Lett.}\ }\textbf {\bibinfo
  {volume} {117}},\ \bibinfo {pages} {136602} (\bibinfo {year}
  {2016})}\BibitemShut {NoStop}%
\bibitem [{\citenamefont {Wang}\ \emph {et~al.}(2016)\citenamefont {Wang},
  \citenamefont {Lu},\ and\ \citenamefont {Shen}}]{PhysRevLett.117.077201}%
  \BibitemOpen
  \bibfield  {author} {\bibinfo {author} {\bibfnamefont {C.~M.}\ \bibnamefont
  {Wang}}, \bibinfo {author} {\bibfnamefont {H.-Z.}\ \bibnamefont {Lu}}, \ and\
  \bibinfo {author} {\bibfnamefont {S.-Q.}\ \bibnamefont {Shen}},\ }\href
  {\doibase 10.1103/PhysRevLett.117.077201} {\bibfield  {journal} {\bibinfo
  {journal} {Phys. Rev. Lett.}\ }\textbf {\bibinfo {volume} {117}},\ \bibinfo
  {pages} {077201} (\bibinfo {year} {2016})}\BibitemShut {NoStop}%
\bibitem [{\citenamefont {Zyuzin}\ and\ \citenamefont
  {Burkov}(2012)}]{PhysRevB.86.115133}%
  \BibitemOpen
  \bibfield  {author} {\bibinfo {author} {\bibfnamefont {A.~A.}\ \bibnamefont
  {Zyuzin}}\ and\ \bibinfo {author} {\bibfnamefont {A.~A.}\ \bibnamefont
  {Burkov}},\ }\href {\doibase 10.1103/PhysRevB.86.115133} {\bibfield
  {journal} {\bibinfo  {journal} {Phys. Rev. B}\ }\textbf {\bibinfo {volume}
  {86}},\ \bibinfo {pages} {115133} (\bibinfo {year} {2012})}\BibitemShut
  {NoStop}%
\bibitem [{\citenamefont {Andreev}\ and\ \citenamefont
  {Spivak}(2018)}]{PhysRevLett.120.026601}%
  \BibitemOpen
  \bibfield  {author} {\bibinfo {author} {\bibfnamefont {A.~V.}\ \bibnamefont
  {Andreev}}\ and\ \bibinfo {author} {\bibfnamefont {B.~Z.}\ \bibnamefont
  {Spivak}},\ }\href {\doibase 10.1103/PhysRevLett.120.026601} {\bibfield
  {journal} {\bibinfo  {journal} {Phys. Rev. Lett.}\ }\textbf {\bibinfo
  {volume} {120}},\ \bibinfo {pages} {026601} (\bibinfo {year}
  {2018})}\BibitemShut {NoStop}%
\bibitem [{\citenamefont {Son}\ and\ \citenamefont
  {Spivak}(2013)}]{PhysRevB.88.104412}%
  \BibitemOpen
  \bibfield  {author} {\bibinfo {author} {\bibfnamefont {D.~T.}\ \bibnamefont
  {Son}}\ and\ \bibinfo {author} {\bibfnamefont {B.~Z.}\ \bibnamefont
  {Spivak}},\ }\href {\doibase 10.1103/PhysRevB.88.104412} {\bibfield
  {journal} {\bibinfo  {journal} {Phys. Rev. B}\ }\textbf {\bibinfo {volume}
  {88}},\ \bibinfo {pages} {104412} (\bibinfo {year} {2013})}\BibitemShut
  {NoStop}%
\bibitem [{\citenamefont {Xiao}\ \emph {et~al.}(2017)\citenamefont {Xiao},
  \citenamefont {Law},\ and\ \citenamefont {Lee}}]{PhysRevB.96.165101}%
  \BibitemOpen
  \bibfield  {author} {\bibinfo {author} {\bibfnamefont {X.}~\bibnamefont
  {Xiao}}, \bibinfo {author} {\bibfnamefont {K.~T.}\ \bibnamefont {Law}}, \
  and\ \bibinfo {author} {\bibfnamefont {P.~A.}\ \bibnamefont {Lee}},\ }\href
  {\doibase 10.1103/PhysRevB.96.165101} {\bibfield  {journal} {\bibinfo
  {journal} {Phys. Rev. B}\ }\textbf {\bibinfo {volume} {96}},\ \bibinfo
  {pages} {165101} (\bibinfo {year} {2017})}\BibitemShut {NoStop}%
\bibitem [{\citenamefont {Hershfield}\ and\ \citenamefont
  {Ambegaokar}(1986)}]{PhysRevB.34.2147}%
  \BibitemOpen
  \bibfield  {author} {\bibinfo {author} {\bibfnamefont {S.}~\bibnamefont
  {Hershfield}}\ and\ \bibinfo {author} {\bibfnamefont {V.}~\bibnamefont
  {Ambegaokar}},\ }\href {\doibase 10.1103/PhysRevB.34.2147} {\bibfield
  {journal} {\bibinfo  {journal} {Phys. Rev. B}\ }\textbf {\bibinfo {volume}
  {34}},\ \bibinfo {pages} {2147} (\bibinfo {year} {1986})}\BibitemShut
  {NoStop}%
\bibitem [{\citenamefont {Kim}\ \emph {et~al.}(2014)\citenamefont {Kim},
  \citenamefont {Kim},\ and\ \citenamefont {Sasaki}}]{PhysRevB.89.195137}%
  \BibitemOpen
  \bibfield  {author} {\bibinfo {author} {\bibfnamefont {K.-S.}\ \bibnamefont
  {Kim}}, \bibinfo {author} {\bibfnamefont {H.-J.}\ \bibnamefont {Kim}}, \ and\
  \bibinfo {author} {\bibfnamefont {M.}~\bibnamefont {Sasaki}},\ }\href
  {\doibase 10.1103/PhysRevB.89.195137} {\bibfield  {journal} {\bibinfo
  {journal} {Phys. Rev. B}\ }\textbf {\bibinfo {volume} {89}},\ \bibinfo
  {pages} {195137} (\bibinfo {year} {2014})}\BibitemShut {NoStop}%
\bibitem [{\citenamefont {Das}\ and\ \citenamefont
  {Agarwal}(2019)}]{PhysRevB.99.085405}%
  \BibitemOpen
  \bibfield  {author} {\bibinfo {author} {\bibfnamefont {K.}~\bibnamefont
  {Das}}\ and\ \bibinfo {author} {\bibfnamefont {A.}~\bibnamefont {Agarwal}},\
  }\href {\doibase 10.1103/PhysRevB.99.085405} {\bibfield  {journal} {\bibinfo
  {journal} {Phys. Rev. B}\ }\textbf {\bibinfo {volume} {99}},\ \bibinfo
  {pages} {085405} (\bibinfo {year} {2019})}\BibitemShut {NoStop}%
\bibitem [{\citenamefont {Deng}\ \emph
  {et~al.}(2019{\natexlab{b}})\citenamefont {Deng}, \citenamefont {Duan},
  \citenamefont {Luo}, \citenamefont {Deng}, \citenamefont {Wang},\ and\
  \citenamefont {Sheng}}]{PhysRevB.99.165146}%
  \BibitemOpen
  \bibfield  {author} {\bibinfo {author} {\bibfnamefont {M.-X.}\ \bibnamefont
  {Deng}}, \bibinfo {author} {\bibfnamefont {H.-J.}\ \bibnamefont {Duan}},
  \bibinfo {author} {\bibfnamefont {W.}~\bibnamefont {Luo}}, \bibinfo {author}
  {\bibfnamefont {W.~Y.}\ \bibnamefont {Deng}}, \bibinfo {author}
  {\bibfnamefont {R.-Q.}\ \bibnamefont {Wang}}, \ and\ \bibinfo {author}
  {\bibfnamefont {L.}~\bibnamefont {Sheng}},\ }\href {\doibase
  10.1103/PhysRevB.99.165146} {\bibfield  {journal} {\bibinfo  {journal} {Phys.
  Rev. B}\ }\textbf {\bibinfo {volume} {99}},\ \bibinfo {pages} {165146}
  (\bibinfo {year} {2019}{\natexlab{b}})}\BibitemShut {NoStop}%
\end{thebibliography}%

\end{document}